\documentstyle[11pt,mrs2001,epsfig]{article}
\begin{document}
\title{THE NATURE OF THE FAINT RADIO POPULATION}

\author{I. PRANDONI$^{1,2}$, L. GREGORINI$^{1,2}$, P. PARMA$^2$, H.R. de 
RUITER$^{3,2}$, G. VETTOLANI$^2$, M. WIERINGA$^4$, R. EKERS$^4$}
\affil{$^1$Universit\`a degli Studi di Bologna, Bologna, Italy}
\affil{$^2$Istituto di Radioastronomia del CNR, Bologna, Italy}  
\affil{$^3$Osservatorio Astronomico di Bologna, Bologna, Italy}  
\affil{$^4$Australia Telescope National Facility, Sydney, Australia}

\begin{abstract}
We present the results obtained so far from the optical follow-up of the 
ATESP sub-mJy radio survey. The ATESP results are then compared with the ones 
obtained from other deep radio samples and we show how the existing 
discrepancies about the nature of the faint radio population 
can be explained in terms of selection effects.
\end{abstract}

\section{Introduction}

Radio source counts derived from deep 1.4 GHz surveys 
show a sudden steepening below 1 mJy (e.g. \cite{Windhorst90}, 
\cite{Prandoni01a}), which is interpreted as the result of
the emergence of a new population of radio sources different in nature from
the ones which dominate at 
higher flux densities (e.g. classical radio galaxies and QSOs). Nevertheless,
despite the large observational efforts, the physical and evolutionary 
properties of this population are still poorly understood. \\
Today we know that the faint radio population is a mixture of several types 
of objects 
(faint AGNs, normal spirals and ellipticals, starburst galaxies), but the 
relative importance of the different classes is still debated and very little 
is known about their redshift distribution and luminosity properties. \\
The cause of it is twofold. First, faint radio samples are
small, with sky coverages typically going from a fraction to a few square 
degrees. The statistics available is therefore quite poor. Secondly and more 
importantly, the optical follow-up of these samples is strongly incomplete. 
Typically $50\% - 60\%$ of the radio sources are identified on optical images,
while only $\sim 20\%$ have spectroscopic information. Exceptions are the 
$\mu$Jy sample taken on the HDF North (see \cite{Richards}) where the 
identification rate is $80\%$ and the Marano Field (MF) sub-mJy sample 
(see \cite{Gruppioni}) with spectroscopy available for $45\%$ of the 
sources. In both cases, though, the numbers involved are very small. \\
This means that the results drawn from radio-optical studies of the 
faint radio population are based on limited optical follow-up and biased 
by the fact that only the optically brightest sources have spectral 
information available. It is therefore clear that larger deep radio samples 
with possibly complete optical follow-up are strongly needed in order to 
fully assess the nature and evolution of the faint radio population. 

\section{The ATESP Radio Survey and its Optical Follow-up}

With this in mind, we took advantage of the mosaic observing mode 
capability of the Australia Telescope Compact Array (ATCA) to deeply image
at 1.4 GHz the entire region previously covered by the ESO Slice Project (ESP)
redshift survey (see \cite{Vettolani} for a full description). \\
We have produced 16 radio mosaics with uniform noise of $\sim 79$ $\mu$Jy and 
spatial resolution of $\sim 8\arcsec \times 14\arcsec$ (see \cite{Prandoni00a}
for details). The sky coverage of
the ATESP survey is 26 square degrees, i.e. one order of magnitude larger than
any other previous sub-mJy sample. 
From the radio images we have extracted a catalogue of 2967 radio sources, 
complete down to $S\sim 0.5$ mJy (see \cite{Prandoni00b}). \\
The ESP redshift survey (\cite{Vettolani}) 
provided us with spectroscopy information for a sample of 3342 galaxies 
complete down to $b_j \sim 19.4$. Such data allowed us to identify about 10\%
of the ATESP radio sources. The typical depth of the ESP survey ($0<z<0.3$) 
is $z\sim 0.1$. Such sample is therefore very well suited to assess the 
{\it local} radio-optical properties of the faint radio population.
In Fig.~1 we show the bivariate radio luminosity function  
derived from the ESP galaxies, as a function of the well-known 
radio-to-optical ratio ($R$). $R = S \cdot 10^{0.4(m-12.5)}$
where $S$ is the 1.4 GHz radio flux in mJy and $m$ is the $b_j$ optical 
magnitude (see \cite{Condon}). 
The radio-to-optical ratio is a measure of the radio excess in a 
galaxy of given optical luminosity. As shown in the figure, emission
line galaxies (typically associated with spiral galaxies) are characterized by 
small radio-to-optical ratios and their luminosity function steeply increases 
below a reference value of $R\simeq 250$ (or $\log{R}\simeq 2.4$). On the 
other hand, absorption systems (typically associated with elliptical and S0 
galaxies) have a flatter luminosity function and their 
contribution to the overall population increases going to higher 
radio-to-optical ratios, becoming the dominant class at very high $R$-values.\\
This result implies that radio emission associated to star-forming regions in 
disk galaxies is weaker than radio emission triggered by nuclear 
activity in early-type galaxies. \\

\begin{figure}[t]
\plotone{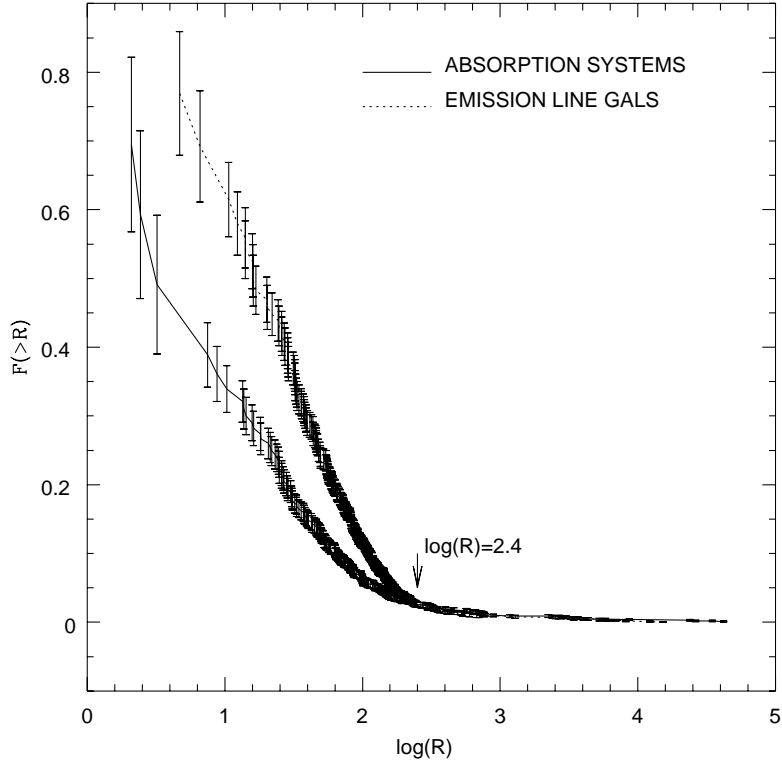}{12cm}
\caption{Local radio-optical luminosity function $F(R)$ expressed in terms of 
the radio-to-optical ratio $R$. $F(R)$ has been derived separately for ESP
galaxies with (dotted) and without (solid) emission lines in their 
optical spectra. Also indicated is the value of $R$ below which the emission
line galaxy population dominates ($\log{R}\simeq 2.4$).}
\end{figure}

\begin{figure}[t]
\plotone{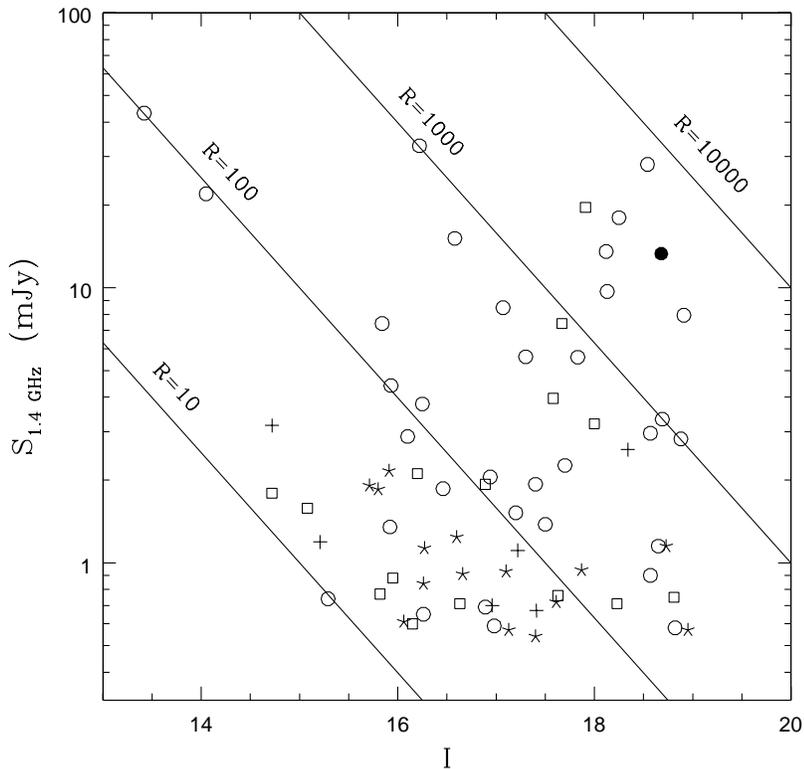}{12cm}
\caption{1.4 GHz flux density (in mJy) versus I magnitude for the 70 ATESP-EIS 
sources with $I<19$. Symbols represent different spectral classes: 
Early type (empty circles), Late type (empty squares), SB + post-SB 
(stars) and AGN (crosses). The filled circle indicates the object with no 
spectroscopy available. Solid lines represent constant radio-to-optical ratios
(here computed using the I magnitude).}
\end{figure}

\noindent
To get optical information about ATESP sources at higher redshifts, we have 
focused our attention on a sub-region of 3 sq. degrees which overlaps with the ESO 
Imaging Survey (EIS Patch A). In this sub-region optical images and catalogues
complete down to $I\sim 22.5$ have been made publicly available in 1998 
(\cite{Nonino}). Such data allowed us to identify 60\% ($219/386$) of the 
ATESP sources in this area (see \cite{Prandoni01b} for details). 
We have then used the ESO 3.6 m telescope to get spectroscopy of the 
identified sources in the ATESP-EIS region. So far we have obtained spectra
for 69 of the 70 galaxies brighter than $I=19$. This spectroscopic sample can
therefore considered complete down to this limiting magnitude. The median 
redshift of the sample is $z\sim 0.2$ and the good quality of our spectroscopy
allowed us to reliably classify all the observed objects. \\
By inspecting the ATESP radio source composition as a function of flux,
we find that the AGN contribution  does not
significantly change going from mJy to sub-mJy fluxes ($8-9\%$). On the other
hand early type galaxies largely dominate (60\%) the mJy
population, while star-formation processes become important at sub--mJy
fluxes: SB and post-SB galaxies go from 13\% at $S\geq 1$ mJy to 39\% at
$S<1$ mJy. Nevertheless, at sub--mJy fluxes, early type galaxies still 
constitute a significant fraction (25\%) of the whole population 
(\cite{Prandoni01b}). A result 
which was already noticed in the Marano Field (\cite{Gruppioni}) and in the 
Phoenix Deep Field (\cite{Georgakakis}). 
The change with flux of the source composition is evident 
in the flux-magnitude diagram shown in Fig.~2. Early type galaxies (empty 
circles)
are the dominant population at higher flux densities, while the contribution 
of star-forming galaxies (stars) becomes increasingly important 
going to sub-mJy fluxes. At a closer inspection, though, we notice that, as
found in our local sample, star-forming galaxies preferentially
populate the region of the plot corresponding to low radio-to-optical ratios, 
while the opposite is true for early type galaxies. \\
This suggests that the main parameter that regulates the faint radio source
composition is not the radio flux alone, but rather the radio-to-optical 
ratio, i.e. a combination of the radio flux and the optical magnitude. 
Under this hypothesis we expect that at fainter magnitudes the ATESP radio
sources will be mostly early type galaxies and AGNs (supporting the results
of \cite{Gruppioni} in the Marano Field), while at a given 
limiting magnitude, fainter radio samples will be more sensitive to the 
star-forming galaxy population (as supported by the HDF $\mu$Jy sample 
\cite{Richards}). 

\section{Comparison with Other Faint Radio Samples}

\begin{figure}[t]
\plotone{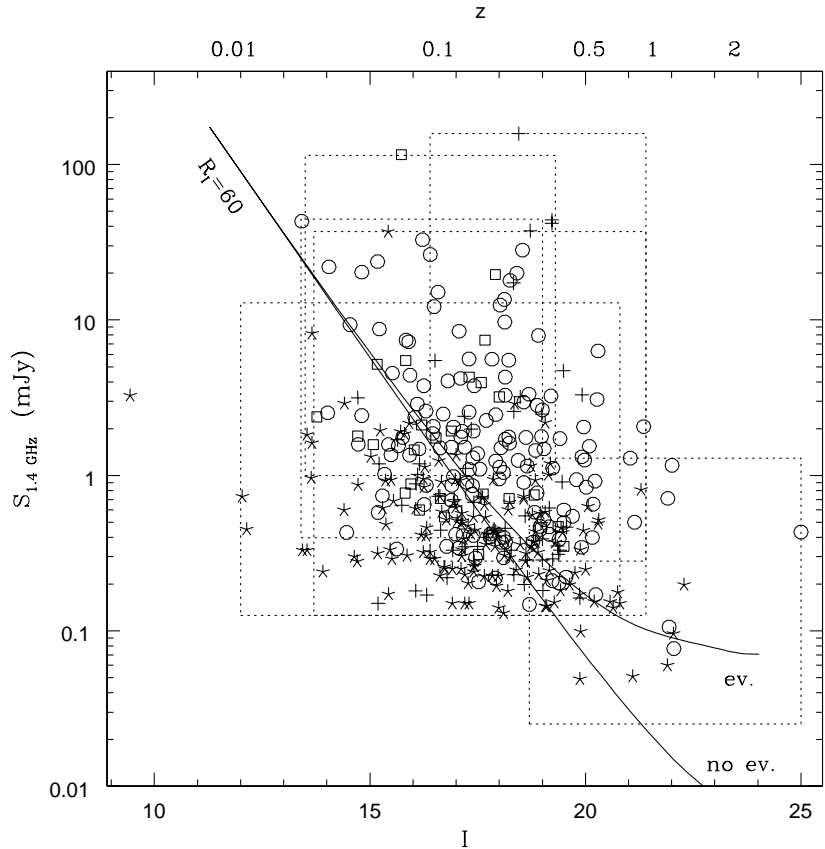}{12cm}
\caption{1.4 GHz flux density versus I magnitude for sources belonging to 
several different faint radio samples. Symbols are as in Fig.~2. The dotted 
boxes represent the region of the diagram probed by each of the plotted 
samples. Solid lines: reference $R$-value below which star-forming galaxies 
are expected to dominate (with and without evolution included). 
Top axis: indicative redshift values. }
\end{figure}

The picture outlined above needs verification on a larger statistical base, 
getting spectroscopy down to much fainter magnitudes and including complete
radio-optical analysis of large $\mu$Jy samples. Nevertheless, a first 
verification can be done using the existing data in the literature.\\
In Fig.~3 we show a flux-magnitude diagram, with 
all the mJy, sub-mJy and $\mu$Jy radio samples available so far included
(FIRST \cite{Magliocchetti}, ATESP-EIS \cite{Prandoni01b}, MF 
\cite{Gruppioni}, PDF \cite{Georgakakis}, B93 \cite{Benn}, H00 
\cite{Haarsma}).  
Only the sources with measured redshift and reliable classification are 
plotted. 
The lines represent the reference radio-to-optical value, below which 
the star-forming population is expected to dominate. This value
has been locally normalized to the value found in our ESP sample (see Fig.~1):
assuming $b_j-I \sim 1.6$, we have that $R_{b_j}\simeq 250$ translates to 
$R_I \simeq 60$. \\
The samples drawn are affected by several incompleteness and selection 
biases and therefore no strong statement can be drawn from this plot.
Nevertheless it is worth noticing that the distribution of both the early 
type (empty circles) and the star-forming galaxies (stars) follows the 
expected one, once we include some standard form of evolution for both the 
radio ($L(z) \sim (1+z)^3$) and the optical (\cite{Poggianti}) 
luminosity in drawing the reference $R$ line. \\
Very interestingly, this picture provides a very natural way to explain and 
reconcile the existing discrepant results about the relative contribution of 
star-forming and early type galaxies to the faint radio population. 
In Fig.~4 we show the region of the flux-magnitude diagram probed by 
each of the samples drawn in Fig.~3. In each panel, the solid line represents 
the reference $R$-value shown in Fig.~3 (the one with evolution included)
and separates the locus dominated by 
early-type galaxies (above) from the locus dominated by 
star-forming galaxies (below). As expected the fraction of early 
type galaxies found in each sample increases clockwise, going from 10\% in 
the B93 sample, which is the less sensitive to the locus dominated by the 
early type population, to $>50\%$ in the MF, which is the most sensitive to 
early type galaxies. Correspondingly, the fraction of star-forming galaxies 
decreases clockwise from $>50\%$ in the B93 sample to $\sim 30\%$ in the MF.\\
This qualitative analysis seems very promising in giving a unique 
picture of the faint radio population, but more quantitative studies are 
needed to verify this scenario. To this respect very helpful will be 
the analysis of a 1~sq.~degr. region where the ATESP survey overlaps 
with a very deep ($I\sim 26$) multicolor (UBVRI) ESO survey which is 
currently under way. 

\begin{figure}[t]
\plotone{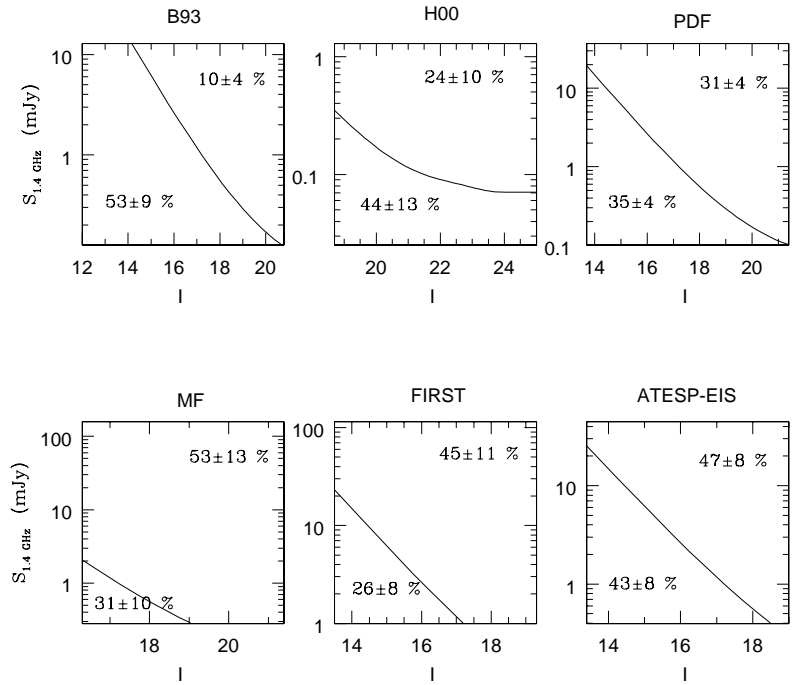}{11cm}
\caption{Flux-magnitude diagrams for the samples plotted in Fig.~3. 
The solid line in each plot represents the {\it evolved} reference $R$-value 
shown in Fig.~3.
Also indicated are the fraction of early type 
(top right) and star-forming (bottom left) galaxies found in each sample. }
\end{figure}

\vfill
\end{document}